# Bell inequality and complementarity loophole

Marek Czachor*

*Katedra Fizyki Teoretycznej i Metod Matematycznych*
*Politechnika Gdańska, ul. Narutowicza 11/12, 80-952 Gdańsk, Poland*

A simple classical, deterministic, local situation violating the Bell inequality is described. The detectors used in the experiment are ideal and the observers who decide which pair of measuring devices to choose for a given pair of particles have free will. The construction uses random variables which are not jointly measurable in a single run of an experiment and the hidden variables have a nonsymmetric probability density. Such random variables are complementary but still fully classical. An assumption that classical random variables cannot satisfy any form of complementarity principle is false, and this is the loophole used in this example. A relationship to the detector inefficiency loophole is discussed.

## I. INTRODUCTION

Consider some system which is characterized by the following properties. It consists of a classical source which emits pairs of classical particles which are clasically correlated. There are two observers, Alice and Bob, who measure binary random variables, say $A$ and $A'$ for Alice, and $B$ and $B'$ for Bob. Alice and Bob use perfect detectors and all pairs emitted by the source are detected in coincidence. Neither Alice and Bob nor the particles themselves communicate and the observers have a possibility of deciding at the very last moment which configuration of $A$ and $A'$, and $B$ and $B'$ to measure. Therefore not only Alice (Bob) does not know what are the settings of $B$ and $B'$ ($A$ and $A'$) for a given pair, but also the particle that propagates towards Bob (Alice) cannot predict which detector will be finally chosen for the measuremnt of the binary random variable. Alice and Bob have free will which means they either decide on their own how to configure the measuring devices, or leave this to a random generator. Once this choice is made the result is completely determined by the classical state of the particles.

I believe that after such a characterization of what is going to happen most of the readers will conclude that the Bell inequality cannot be violated in this experiment. However, in what follows I will give an example of a simple classical system (pairs of billard balls) which satisfy all those requirements and yet violate the Bell inequality. There is of course no magic in this result. I simply take advantage of another loophole which exists in the proof of the Bell theorem: Although the particles and the observables are classical, the experiment is devised in such a way that on a given pair one can measure *either* $A$ and $B$, *or* $A$ and $B'$, *or* $A'$ and $B$, *or* $A'$ and $B'$. Once we have a result, say $A = +1$, it makes no sense to compare it to the result of $A'$, because there is no such result for this pair! We have a kind of classical complementarity here and it turns out to be sufficient for the violation. Before I explain how the model works let me therefore begin with a few comments on complementarity principle in a context of hidden variables.

## II. COMPLEMENTARITY OF RANDOM VARIABLES

There exists a prejudice that the very idea of hidden variables contradicts the notion of *complementarity*. It is argued that once we know a hidden variable state $\lambda$ we know — by definition — also the values of all random variables $A(\lambda)$, $A'(\lambda)$, etc., even if their corresponding quantum counterparts (observables) $\hat A$, $\hat A'$, are not simultaneously measurable. In particular, it is assumed that at a hidden variables level one can safely consider expressions such as $A(\lambda) + A'(\lambda)$ even though the two random variables represent non-orthogonal linear polarizations of a single photon. The Bell inequality [1]

$$|\mathcal{B}_{A,A',B,B'}| \leq 2$$

is obtained if one averages the "Bell random variable"

$$\mathcal{B}_{A,A',B,B'}(\lambda) = \big[A(\lambda) + A'(\lambda)\big]B(\lambda) + \big[A(\lambda) - A'(\lambda)\big]B'(\lambda).$$

That something may be fundamentally wrong with random variables of the form $A + A'$ is illustrated by the simple classical Aerts model discussed in [2,3]. The system he considered consisted of a mass $m$ placed on a circle and a measuring device composed of two masses $m_1$, $m_2$, satisfying $m_1 + m_2 = 1$ and placed antipodally on the circle. A measuremnt gives "+1" if the Newton force between $m$ and $m_1$ is greater than this between $m$ and $m_2$. In such a case the mass $m$ falls on the location point of $m_1$. To make another measurement one leaves $m$ at the point of its arrival, removes the masses $m_1$ and $m_2$ and puts another, similar pair $m'_1$, $m'_2$ of masses tilted by some $\theta$ with respect to the previous one. The conditional probability of "first +1 and second +1" for two such measurements made *one after another* is $\cos^2(\theta/2)$ which is typical of spin-1/2 particles. Since the probability is of a Malusian type there exists a classical Bell-type inequality which can be derived from the classical Bayes rule but which is not satisfied by the system of masses on the circle. The hidden variables are here $\alpha \in [0, 2\pi)$ (the position of $m$) and $m_1 \in [0, 1]$. After the first measurement we know $\alpha$ and the remaining probability follows from our lack of knowledge about $m'_1$.

Obviously, a knowledge of both $\alpha$ and $m_1$ completely determines a result of a measurement: $A(\lambda) = A(\alpha, m_1)$. It makes perfect sense to consider a measurement of *first* $A$ and *next* $A'$. The conditional probabilities for such a case are the spin-1/2 ones. It is also fully justified to ask the *counterfactual* question: "What are the probabilities of finding such $\alpha$ and $m_1$ that we would have obtained $A(\alpha, m_1) = +1$ if we had decided to measure $A$, and $A'(\alpha, m_1) = +1$ had we decided to measure $A'$?" This problem was discussed in detail in [3] and it was shown that the counterfactual probabilities are *not* the spin-1/2 ones and *do* satisfy the Bell inequality. However, these probabilities are not those which we measure in a real experiment. The point is that the measurement involves a *change of state* (an initial $\alpha$ changes as the mass $m$ moves in the direction of either $m_1$ or $m_2$). Therefore it simply makes no sense to consider *simultaneously* two such results in an actual experiment because the mass $m$ cannot move in two different directions simultaneously. The two measurements are *complementary* and this is why it is *logically impossible* to derive the Bell inequality for an actual experiment even though the situation is classical. On the other hand the counterfactual *Gedankenexperiment* leads to the Bell inequality but the resulting probabilities are not measurable *in principle*.

I find this example very instructive because it shows that the high-school axioms of probability, which are taken for granted in all "reasonable" hidden variable models, may not be satisfied even in macroscopic and fully classical situations. However, the Aerts example is in fact two-edged. If one tries to construct a kind of a singlet state situation involving two cirles one obtains the correct probabilities provided a measurement on one circle polarizes the mass on the other one, and this implies nonlocality. So the result is in a sense trivial: One can violate the inequality in a hidden variables model but for the price of locality. This was clearly stated already in the 1964 Bell paper. Actually, Bell proposed in this paper a formal model which gave the singlet state (or equivalently spin-1/2) probabilities and used implicitly the trick with two mutually influencing measurements but acting at a distance. It follows that the Bell-type nonlocality is, from a hidden variables perspective, a kind of complementarity-at-a-distance.

The examples proposed by Aerts and Bell *suggest* that complementarity can be formulated in hidden variables terms provided one introduces mutually disturbing measurements. However, the statement we have proved works, strictly speaking, the other way around: Once we have measurements that influence each other, the corresponding random variables may be thought of as mutually complementary. We have *not* proved that there do not exist other representations of complementarity in the framework of hidden variables. In fact, the next section discusses another such case. It will be shown that a *local complementarity* may be also sufficient for a violation of the Bell inequality. This is what I call the complementarity loophole. The model I describe is macroscopic, local, fully deterministic, involves 100% efficient detectors, can be made by any kid at home, and yet maximally violates the Bell inequality. It is similar to the situation I once discussed in [4].

### III. THE MODEL

We consider two systems, say billard balls, which are initially somehow connected and form a "molecule". Such a "molecule" is placed at random in a room which has eight square holes cut in its floor. The holes form a chessboard-like pattern shown in Fig. 1. We assume that each "molecule" will finally fall out through one of the holes, and once it reaches the next floor it breaks and separates into two disconnected billard balls which start to roll towards two perpendicular walls of the room. Let us call the walls the "Alice wall" and the "Bob wall". We assume that the room and the balls are devised in such a way that if one of the balls moves towards the Alice wall the other one moves towards the Bob one. Now, once they reach the walls they fall into one of two drawers, say $A$ and $A'$ for the Alice wall, and $B$ and $B'$ for the Bob wall. Each of the drawers contains two boxes: the right one and the left one. Once Alice or Bob hear a sound of a ball falling into her or his drawer they open it and give a result "+1" if they find the ball in the right box and "−1" if they find it in the left one. Each drawer defines a binary random variable: $A$, $A'$ for the Alice wall and $B$, $B'$ for the Bob one.

Mathematically we may model the experiment as follows. The chessboard is the $[0, 4] \times [0, 4]$ square. The falling "molecule" falls at a point with coordinates $(a, b)$. If $a \in [0, 2]$ then Alice finds the ball in $A'$ and obtains "−1" if $a \in [0, 1]$ and "+1" if $a \in (1, 2]$. For $a \in (2, 4]$ Alice finds the ball in $A$ and obtains "−1" if $a \in (2, 3]$ and "+1" if $a \in (3, 4]$. We similarly describe the measurements made by Bob.

We finally add one more element. We assume that both Alice and Bob can freely exchange the locations of the drawers. So, for example, Alice can cross the possible paths of the balls which move towards her drawers according to the scheme shown at Fig. 2. She does it without knowing whether a given ball moves towards $A$ or $A'$. She makes the decision either according to her free will or leaves it to a random generator. In either case the probability that the drawers will be exchanged is $p_A$ for Alice and $p_B$ for Bob. So neither Alice nor Bob can know for sure which experiment is made by her or his partner.

Let us note that the arrangement is in a sense similar to the experiment of Aspect [5,6] where the devices remain in fixed positions but photons from the singlet state pair choose the devices at random. In the Aspect-type experiment the choice depends on whether the photon is reflected or transmitted at an optical switch and this, in principle, depends on both the (hidden variable) state of the photon and this of the switch.

### A. Locality

From the description given above it follows that Alice and Bob do not communicate. The results they obtain depend on the initial correlation (state) of the "molecule", that is the coordinates of the point on the floor where the "molecule" splits into two balls, and the local actions they undertake. Therefore the situation is local.

### B. Determinism

The system is deterministic. The determinism does not necessarily apply to the "switch" since a free will of Alice and Bob can change its state. A "particle" does not "know" which "dectector" will be chosen since the "switch" can be changed to another state at any time.

### C. Detection efficiency

To determine the detector efficiency we compare the number of detections in coincidence to those where only one of the detectors "clicks". Here whenever Alice's $A$ or $A'$ "clicks" then either $B$ or $B'$ does the same, and vice versa. We conclude that the detectors are perfect.

### D. Averages

Averages of the "one arm" random variables vanish: $\langle A \rangle = \langle A' \rangle = \langle B \rangle = \langle B' \rangle = 0$ but the Bell average depends on the type of actions undertaken by Alice and Bob:

$$\langle AB \rangle + \langle AB' \rangle + \langle A'B \rangle - \langle A'B' \rangle = 4p_A p_B. \quad (1)$$

The Bell inequality is violated for $p_A p_B > 0.5$.

### E. Perfect correlations

The averages occuring in (1) are $\pm 1$ if $p_A = p_B = 1$ so that the inequality can be violated even for perfect correlations. This makes our "state" in a sense similar to the Greenberger-Horn-Zeilinger (GHZ) one [7].

### F. Complementarity

Similarly to real Bell-type experiments Alice and Bob measure, on a single pair, only either $A$ and $B$, or $A$ and $B'$, or $A'$ and $B$, or $A'$ and $B'$. Alice cannot measure simultaneously $A$ and $A'$. This is simply logically impossible because the hidden variable $a$ (the coordinate of the "molecule") cannot simultaneously satisfy $a \leq 1$ and $a > 1$.

### G. Symmetry

A transition between $A$ and $A'$, and $B$ and $B'$ is obtained by a shift of the domains of the random variables by 2 to the left or to the right. The initial statistical state of the pair (the chessboard) is not invariant under this operation. This element is technically responsible for the violation (one cannot change the variables under integrals and do not change the probability density).

### H. Symmetry vs. complementarity

The above discussion shows that alternative measurements $A$ and $A'$ may be called complementary if a transformation $A \to A'$ applied to random variables is not a symmetry of the probability density of the hidden variables. So an additional care is needed if one tries to describe in hidden variables terms states which do not possess symmetries. This remark applies in particular to the theorems of Home and Selleri [8], and Gisin and Peres [9], which use general entangled states, and to the GHZ states which are not rotationally invariant.

### I. Free will of Alice and Bob

Alice and Bob are free to manipulate with the detectors. The degree of their freedom is measured by $p_A$ and $p_B$. Their different actions result in different averages they obtain. In fact, the more active they are the more violation is obtained. This clearly distinguishes our example from the situation discussed by Brans in [10] where the choice of detectors is pre-determined since his universe is totally determinstic.

### J. Data collecting procedures

The experiment with the drawers formally resembles the Aspect-type experiment with the optical switch. There exists another class of experiments [11–15] where the detectors remain in fixed positions during a given run. Alice and Bob may insist on collecting only the data from $A$ and $B$ in a given run of an experiment. In such a case there will be a fraction of particles, namely those that move towards $A'$ and $B'$, that will not be detected. This resembles the problem of detector inefficiency loophole [16]. In experiments one usually tries to collect as many pairs as possible by using waveguides or mirrors to direct those propagating in the "wrong" directions into the fixed detectors so that no data will be lost. An alternative procedure is to surround a source with detectors in a way guaranteeing that no particles will escape. The latter case corresponds to the procedure used in our experiment. Obviously, we could use also the other procedure by directing the balls which move towards the "wrong" drawers into the "right" ones. It is intersting that in such a case the Bell inequality would not be violated. The reason is that in a formal description of the experiment we would have to extend domains of all random variables appearing in (1) onto the whole space of hidden variables and then the standard trick leading to the Bell inequality would work. The two procedures are therefore, in general, physically inequivalent although the opposite is typically assumed.

### K. Alternative choices of detectors

The alternative choices of our "detectors" correspond physically to a situation where instead of rotating a single polarizer from $A$ to $A'$ we have two polarizers $A$ and $A'$ located in different places and we exchange their locations. In general if the state one considers is nonsymmetric one may expect a correlation between the direction of propagation and the internal state which is measured by Alice or Bob. This is what happens here albeit quite trivially.

### L. Quantum description

Any classical model allows also for a quantum description. To find it we can introduce, instead of the billard balls, pairs of photons or any other quantum particles which are emitted by a source which is classically fluctuating in the way determined by the chessboard from Fig. 1. To each drawer one can associate a projector in position space which corresponds to a result, say, "a photon is detected in the left box in the drawer $A$". Having these projectors we can construct a density matrix which describes the source. The observables measured by Alice and Bob can be also constructed in terms of the projectors. Now a warning. Of course any projector has eigenvalues 1 and 0. The states corresponding to a detection in $A'$ are in a space orthogonal to those corresponding to a detection in $A$. Therefore one can say that a result $A' = +1$ is simultaneously equivalent to $A = 0$. Therefore $A$ has three results, $\pm 1$ and 0, and the Bell inequality should be derived for random variables with three results. In a sense this is true and this ambiguity was discussed in some detail in [4]. In more recent literature it is called a "Kolmogorovian censorship" problem [17]. Still notice that if we rull out on this basis our problem as trivial, we have to equally treat the celebrated Aspect experiment with the optical switch. Indeed, a detection in his analyzer $A$ means no detection in $A'$ and the whole argument can be repeated. This also shows that an addition of the value 0 would lead to averages which are not directly related to the detector counts. The masured probabilities are, in fact, the conditional ones and the condition is "provided the particle is detected in $A$".

### IV. CONLUSIONS

I have presented an example of a classical system which is fully deterministic, local, uses perfect detectors, formally resembles the Aspect experiment with optical switch, and yet violates the Bell inequality. The example uses a version of a hidden variables complementarity principle. This is one of the possible versions of such a principle. The other is the Bell-Aerts one where the complementarity is related to mutually disturbing measurements. The Bell nonlocality is a kind of a complementarity-at-a-distance. The complementarity I use is local. The example is rather primitive and perhaps somewhat artificial, but this is a result of my inability of inventing a more sophisticated hidden-variables form of local complementarity. Nevertheless the fact that I could not find anything more convincing does not mean that this is impossible in principle. It seems that one of the conceptual difficulties lies in a sort of a mental block inclining us to use the Komogorovian probability calculus which is naturally related to characteristic functions, i.e. commuting projectors. The commuting projectors are inherently related to situations where no complementarity occurs and do not correctly work even in the classical situation found by Aerts. So complementarity is *not*, in general, a quantum property and there is no reason to believe that it should vanish at levels deeper than the quantum one.

FIG. 1. The white squares denote the holes cut in the floor of the upper room.

FIG. 2. Alice has free will. A particle that moves towards the drawer $A$ ($A'$) can be shifted with probability $p_A$ to $A'$ ($A$). Bob can do similar operations with the probability $p_B$. Here the shifts are chosen in such a way that the results $\pm 1$ are not changed. This is not essential but makes calculations simpler. One could consider cases when $\pm 1$ for $A$ goes into $\mp 1$ for $A'$ and so on. To violate the Bell inequality the probabilities of undertaking such actions by Alice or Bob must satisfy, for this particular model, $p_A p_B > 0.5$.

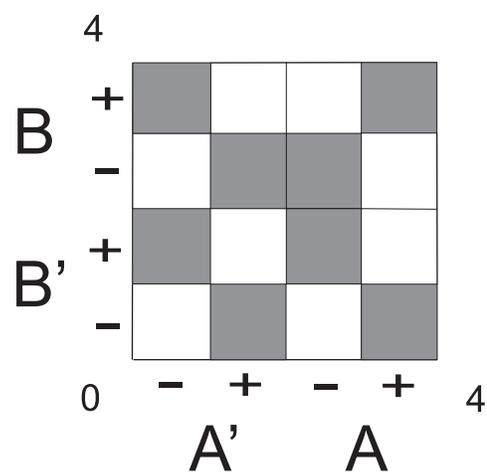

Fig. 1
M. Czachor ,"Bell inequality and complementarity loophole"

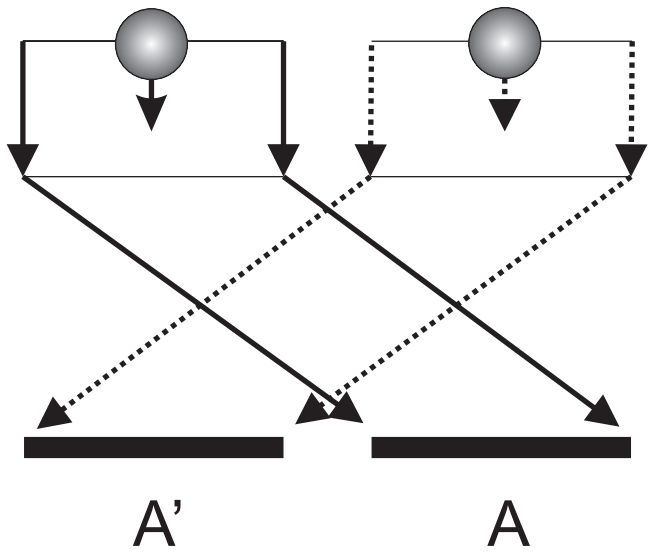

Fig.2
M. Czachor, "Bell inequality and complementarity loophole"